\documentstyle[sprocl]{article}

\def\NPB{{\em Nucl. Phys.} B}

\def\PRL{\em Phys. Rev. Lett.}
\def\PRD{{\em Phys. Rev.} D}

\def\be{\begin{equation}}
\def\ee{\end{equation}}
\def\bea{\begin{eqnarray}}
\def\eea{\end{eqnarray}}

\begin{document}

\title{Signatures of Gauge Mediated Supersymmetry Breaking Models 
at the Tevatron}

\author{B. Dutta}

\address{Institute of Theoretical Physics, University of Oregon,\\ Eugene, 
OR 97403\\E-mail: dutta@oregon.uoregon.edu}


\maketitle\abstracts{ We consider the  scenarios of the GMSB models in which the 
dominant signal for supersymmetry at
the Tevatron are the events having two or three $\tau$ leptons  with
high
$p_T$ accompanied by  large missing transverse energy. This signal is very
different from the multijet or multileptons (involving $e$ and/or $\mu$ only) 
in the usual supergravity models or the  photonic signals in the GMSB models 
(where the lightest neutralino is the next to lightest supersymmetric particle (NLSP)).  
The parameter space  where the lighter
stau is the NLSP  allows 
this possibility. We find that such a signal
could be observable at the Tevatron Run II.\\OITS-355}

\section{Introduction}
At the Tevatron, the dominant processes that give rise to multi $\tau$ signal in the 
GMSB models are the
productions of chargino pairs ($\chi^+\chi^-$) and the chargino and the second
neutralino pairs ($\chi^\pm\chi^0_{2}$). ($\sigma_{\chi^{\pm}\chi^0}$, where
$\chi_0$ is the lightest neutralino is very small compared to
$\sigma_{\chi^{\pm}\chi^0_2}$). The chain of decays of
$\chi^\pm$ and $\chi^0_{2}$ lead to the observable high $p_T$ $\tau$'s and the
missing neutrinos and the gravitinos. There are several mass hierarchies of these
superparticles, (involving $\chi^\pm, \,\chi^0_{2,}\,\chi^{0},\,\tilde\nu_l
(l=e,\mu,\tau),\,\tilde l (l=e,\mu \, {\rm and}\, \tilde\tau_1$) leading to the
inclusive high $p_T$ 2$\tau$ or 3$\tau$ final states plus ${\rlap/E}_T$. Since 
these multi $\tau$ signatures appear in the parameter space where $\tilde \tau_1$
is the NLSP, we
first briefly discuss the GMSB parameter space  giving rise to $\tilde \tau_1$
as the NLSP \cite{all}. 

\section {parameter space} In GMSB models, with radiative EW symmetry breaking, all the sparticle masses
and the mixing angles depend on five parameters,
$M,\,\Lambda,\,n,\,\tan\beta,\,{\rm and\, sign\, of\,}
\mu$.
$M$
 is the messenger scale, and $\Lambda$ is related to the SUSY breaking scale. The parameter
n is fixed by the choice of the vector like messenger sector.  The parameter $\tan\beta$ is the
usual ratio of the up ($H_u$) and down ($H_d$) type Higgs VEVs. The parameter
$\mu $ is the coefficient in the bilinear term, $\mu H_uH_d$ in the
superpotential. The constarints coming from
$b\rightarrow s\gamma$ demands $\mu$ to be mostly negative \cite{ddo}. For n=1, 
the parameter space 
where lighter stau is NLSP has large 
$\tan\beta$ ($\ge 25$) with lower
values of
$\Lambda$. For $n\ge2$, $\tilde\tau_1$ is the NLSP even for the low values of
$\tan\beta$ (for example, $\tan\beta\ge 2$), and for $n\ge3, $
$\tilde\tau_1 $ is again naturally the NLSP for most of the parameter space. The
parameter space where $\tilde\tau_1$ is the NLSP  gives rise to our proposed high
$p_T$ $\tau$ signals. 
\section{Productions and signatures}
At the Tevatron, the chargino pair ($\chi^+\chi^-$) production takes place
through the s-channel  Z and $\gamma$ exchange; while the 
$\chi^0_{2}\chi^{\pm}$ production is via the s channel W exchange. Squark
exchange via the t-channel will also contribute to both  processes. Since the
squark masses are large in the GMSB models these contributions are negligible.
The inclusive final states arising from both of processes are either 2$\tau$ or
3$\tau$ with high
$p_T$ plus ${\rlap/E}_T$ (due to the undetected neutrinos and gravitinos). The
2$\tau$ mode will be mostly of opposite sign charges, but a significant
fraction will also have the same sign of charges. The details of the decays for
the
$\chi^{\pm}$ and $\chi^0_{2}$ depend on the hierarchies of the superparticle
masses, which in turn depend on the GMSB parameter space. There are roughly four possible
cases to consider for EW gaugino production \cite{dn}:
\begin{description}
\centering
\item[Case 1:] $m_{\tilde \nu} > M_{\chi_2^0} \ge M_{\chi_1^\pm}
> m_{\tilde e_1, \tilde \mu_1} > M_{\chi_1^0} > m_{\tilde \tau_1}$
\item[Case 2:] $M_{\chi_2^0} \ge M_{\chi_1^\pm} > m_{\tilde \nu}
> M_{\chi_1^0} > m_{\tilde e_1, \tilde \mu_1} > m_{\tilde \tau_1}$
\item[Case 3:] $M_{\chi_2^0} \ge M_{\chi_1^\pm} > m_{\tilde \nu}
> m_{\tilde e_1, \tilde \mu_1} > M_{\chi_1^0} > m_{\tilde \tau_1}$
\item[Case 4:] $m_{\tilde \nu} > M_{\chi_2^0} \ge M_{\chi_1^\pm} 
> M_{\chi_1^0} > m_{\tilde e_1, \tilde \mu_1} > m_{\tilde \tau_1}$
\end{description} 
The parameter space is also restricted to those regions
where $m_{\tilde\tau_1} \ge 70$ GeV. 
Ongoing LEPII analyses are expected to 
establish this bound soon. 
With this restriction, we did not find any examples for
Cases 3 and 4.\\
Case 1: $m_{\tilde \nu} > M_{\chi_2^0} \approx
M_{\chi_1^\pm} > m_{\tilde e_1, \tilde \mu_1} > M_{\chi_1^0} > 
m_{\tilde \tau_1}$\\
Let us consider a point where
$\tan \beta = 20$, $\Lambda = 32$\,TeV, $M = 480$\,TeV, and $n = 2$.
In the case of chargino pair production,
$\chi_1^\pm \rightarrow \tilde\tau_1 \nu_\tau$ is  the dominant
decay mode. Thus in chargino
pair production, two $\tau$ leptons are always produced. 
In the case of $\chi_1^\pm \chi_2^0$ production the main decay
mode of the second heaviest neutralino is 
$\chi_2^0 \rightarrow \tilde\tau_1 \tau$ with a branching ratio 
(BR) of 85.3\%, while
the only other decay modes are $\chi_2^0 \rightarrow \tilde e_1 e$
and $\chi_2^0 \rightarrow \tilde\mu_1 \mu$. Thus the production
probability for three $\tau$ leptons is high at 85.3\% and the three 
$\tau$-jet
rate is correspondingly 27.2\%. In this parameter point, the
 inclusive 3 $\tau$-jet rate for combined $\chi_1^+ \chi_1^-$
and $\chi_1^\pm \chi_2^0$ production is 9.6\,fb and 
the 2 $\tau$-jets cross section is 62.8\,fb at the RUN II.\\
Case 2: $M_{\chi_2^0} \approx M_{\chi_1^\pm} > m_{\tilde \nu}
> M_{\chi_1^0} > m_{\tilde e_1, \tilde \mu_1} > m_{\tilde \tau_1}$\\
Due to the shifting of the sneutrino masses below
that of $\chi_1^\pm$ and $\chi_2^0$ and also to the shifting
of the selectron and smuon masses below the mass of the lightest 
neutralino, there are now many more decay modes for
$\chi_1^\pm$ and $\chi_2^0$. The dominant decay mode of the
lightest chargino is still $\chi_1^\pm \rightarrow  \tilde\tau_1 \nu_\tau$, 
but now the decays to the sneutrinos are also important. 
In fact, the decay to the sneutrinos can have
branching ratios approaching that of the decay to the stau: for
example in a parameter space where $\tan \beta = 15$, $M = 400$\,TeV, $\Lambda = 20$\,TeV,
and $n = 4$, we have 
BR($\chi_1^\pm \rightarrow \tilde\tau_1 \nu_\tau$) = 0.279, while
BR($\chi_1^\pm \rightarrow \tilde\nu_\tau \tau$) = 0.237. 
For the second lightest neutralino, the dominant
decay mode for these examples is 
$\chi_2^0 \rightarrow \tilde\tau_1 \tau$. But,
as with the decays of the lightest chargino, 
here the decays to the sneutrinos
are important. The branching ratio for the decays of the $\chi_2^0$
to the sneutrinos tend to range from 10\% to 20\% each. 
In this parameter point the rate for inclusive production of 
3 $\tau$-jets is 8.3\,fb and
the inclusive 2 $\tau$-jet rate is 26.2\,fb at the RUN II.
\section{Conclusion}

We have considered the phenomenology of GMSB models where the lighter
stau is the NLSP and
decays within the detector. 
For this situation, the dominant SUSY production processes at the 
Tevatron are $\chi_1^{+}\chi_1^{-}$ and $\chi_1^{\pm}\chi^0_2$. Their prompt 
decays lead to events containing $2\tau$ or $3\tau$ with high $p_T$ plus 
large missing transverse energy.
These signals are different from the photonic signals 
that have been investigated in GMSB models and the dilepton and trilepton
signals in the usual supergravity models.
 The missing transverse energy associated with the events 
is quite large 
providing a good trigger for these events. 

\section*{Acknowledgments}
I would like to thank D.J Muller and S. Nandi for collaboration.
This work was 
supported by DOE grant number DE-FG06-854ER-40224.

\end{document}